%% file: main.tex
\PassOptionsToPackage{english}{babel}
\documentclass[aps, reprint,prl,superscriptaddress,nofootinbib]{revtex4-2}
\usepackage{dcolumn}
\usepackage[utf8]{inputenc}
\usepackage{amsmath}
\usepackage{amsfonts}
\usepackage{amssymb}
\usepackage{graphicx}
\usepackage{hyperref}
\usepackage{xcolor}
\usepackage{adjustbox}
\usepackage[toc,page]{appendix}
\usepackage[english,ngerman]{babel}

\newcommand{\SWAP}{\textsc{swap}}

\newcommand{\PSWAP}{\textsc{pswap}}

\newcommand{\ket}[1]{\ensuremath{|{#1}\rangle}}
\newcommand{\bra}[1]{\ensuremath{\langle{#1}|}}

\begin{document}
\selectlanguage{english}
\title{Temporal witnesses of non-classicality in a 
macroscopic biological system}

\author{Giuseppe Di Pietra}
\email{giuseppe.dipietra@physics.ox.ac.uk}
\affiliation{Clarendon Laboratory, University of Oxford, Parks Road, Oxford OX1 3PU, United Kingdom}

\author{Vlatko Vedral}
\affiliation{Clarendon Laboratory, University of Oxford, Parks Road, Oxford OX1 3PU, United Kingdom}

\author{Chiara Marletto}
\affiliation{Clarendon Laboratory, University of Oxford, Parks Road, Oxford OX1 3PU, United Kingdom}

\date{\today}%

\begin{abstract}
Exciton transfer along a bio-polymer is essential for many biological processes, for instance, light harvesting in photosynthetic biosystems. Here we apply a new witness of non-classicality to this phenomenon, to conclude that, if an exciton can mediate the coherent quantum evolution of a photon, then the exciton is non-classical. We then propose a general qubit model for the quantum transfer of an exciton along a bio-polymer chain, also discussing the effects of environmental decoherence. The generality of our results makes them ideal candidates to design new tests of quantum features in complex bio-molecules. 
\end{abstract}
\maketitle

\section{Introduction}
Quantum theory can in principle be applied to any physical system,  \cite{DEWITT, SCH, wigner1995, deutsch_quantum_1985}, regardless of scale. Its principles explain the stability of matter and are indispensable to understanding the nature of molecular bonding and the dynamics of chemical reactions. This fact, that chemistry is fundamentally quantum regardless of scale, inspired the field of quantum biology, \cite{schrodinger_what_1992,abbott_quantum_2008}, which has now been supercharged by the rapid progress in quantum technologies \cite{lambert_quantum_2013,kim_quantum_2021}. 

The key hypotheses of quantum biology are: (1) that non-trivial quantum effects are present in biological systems, such as light-harvesting complexes in photosynthetic bacteria, or the DNA, or mitochondria \cite{marletto_entanglement_2018,lloyd_quantum_2009,dorner_effects_2012}; (2) that quantum effects enhance biological functionalities, for example by aiding energy transfer along a bio-polymer chain \cite{huelga_vibrations_2013}. 

To this day, experimental evidence for both hypotheses is lacking: while there are many possible quantum models for biological systems, it is difficult to make a conclusive case that classical models (e.g., coupled classical harmonic oscillators) cannot describe them too, as probing complex systems to the same accuracy as, for instance, two entangled photons, is not an easy task. Hence it is essential to find witnesses of quantum effects in biological systems, which could inform realistic experimental schemes to test the validity of the above two hypotheses. Ideally, such witnesses should rely on minimal and plausible physical principles. It is unrealistic to expect that loopholes such as the locality one will be closed when dealing with complex systems anytime soon, hence the need to rely on physical principles. 

To make progress on these issues, here we apply a different strategy compared to previously proposed quantum biology tests. We shall use a recently proposed witness of quantum effects \cite{di_pietra_temporal_2023}, to study an exciton on a generic bio-polymer, i.e., a complex biological system made by many interacting subsystems, the monomers. This witness is based on this protocol: first, one interacts with the bio-polymer via a quantum probe (photons in this case); then, by observing how the probe's dynamical evolution is mediated by the exciton on the bio-polymer, one can establish the exciton's degree of non-classicality. The key physical principle we shall assume is the conservation of energy. 
 
We focus on energy transfer via excitons because it is key for several biological processes, e.g., photosynthesis, in different biological systems of different scales, e.g., \textit{Fenna-Matthews-Olson (FMO) complex} or \textit{polydiacetylene}. Furthermore, it was suggested that quantum coherence in energy transfer may be responsible for its high efficiency \cite{engel_evidence_2007,panitchayangkoon_long-lived_2010, lloyd_quantum_2011}. However, while fully quantum models for exciton transfer are available, classical models can equally well describe it or are compatible with the experimental findings on the process -- hence it has been difficult to assess whether it is genuinely quantum \cite{wilkins_why_2015,duan_nature_2017}. 

Here we shall use the witness of non-classicality to rule out a vast set of classical models as possible descriptions of the exciton transfer along a bio-polymer. We shall use a photon field as a quantum probe, to infer quantum features of the exciton on the bio-polymer, and indirectly of the bio-polymer itself. The strength of this approach is that it doesn't require the direct control of the bio-polymer but of the quantum probe only. This makes the argument applicable to all those biological systems that can be described as complex systems made by properly identified monomers.

To investigate the possibility of applying this witness of quantum effects in a noisy environment, we shall also provide a qubit model of the exciton transfer on a chain of qubits, ideally describing a bio-polymer, without resorting to any dynamical assumption for the latter. This ideal model must not be understood as a detailed description of the exciton dynamics on the bio-polymer but as a general, system and scale-independent way of analysing the best conditions to observe the probe's dynamical evolution required by the witness of quantum effects. This model intends to assess whether the idea of applying the witness of non-classicality to quantum biology can be supported by an experiment on a biological system (the bio-polymer). However, a feasibility study of such an experiment would require enriching the model accordingly, which we leave for future work. 
This makes our information-theoretic argument general enough to be used for different systems in other fields of quantum biology.

\section{Temporal witness of non-classicality}

The word ``non-classicality" shall indicate, in our paper, a specific information-theoretic property.
A system $M$ is \textit{non-classical} if it has at least two distinct physical variables that cannot be measured to arbitrarily high accuracy by the same measuring device \cite{marletto_witnessing_2020}. We call these variables ``incompatible", generalising non-commuting variables in quantum theory. 

A \textit{witness of non-classicality} is a protocol to assess whether a physical system $M$ must be described by at least two incompatible variables, by probing that system with a quantum system. The witness relies on a \textit{witnessing task}, which is defined such that if it can be performed by the system $M$, then $M$ must have two incompatible variables under given assumptions. For instance, in entanglement-based witnesses of non-classicality \cite{marletto_witnessing_2020}, the witnessing task is the creation of entanglement between two spatially separated subsystems $Q_1$ and $Q_2$, mediated solely by $M$. This witness has been applied to quantum gravity in \cite{bose_spin_2017, marletto_gravitationally_2017}. 
The advantage of employing a witnessing task to assess the non-classicality of $M$ lies in its implication that both non-commuting variables are essential to induce $Q$’s dynamical evolution. Consequently, a successful witness of non-classicality would rule out all the classical models describing $M$ that respect the specified assumptions: these classical models would rely on a single variable for $M$, which is insufficient to render the witnessing task possible.
 
Here we shall exploit a different witness \cite{di_pietra_temporal_2023}, which can be regarded as the temporal version of the entanglement-based witness. Let us first consider, for simplicity, a system comprising a single quantum probe $Q$ and the system under investigation $M$. The witnessing task for this witness is the \textit{quantum coherent evolution} of $Q$ driven by $M$, under the assumption that a global quantity on $M$ and $Q$ is conserved. If this task can be achieved in an actual experiment, then we can conclude that $M$ is non-classical.

The witness relies on two assumptions: (i) The conservation of a global variable on $M$ and $Q$, which must be a function of a ``classical" variable $Z_M$ pertaining to $M$, e.g., its energy; (ii) The formalism of quantum theory.
  
To illustrate the argument supporting the witness \cite{di_pietra_temporal_2023}, we shall provide proof by contradiction that assumes, without loss of generality, $Q$ to be a qubit, with $Z_Q$ being its computational basis, and $M$ to be a bit, with $Z_M$ being its ``classical" variable. We shall show that the latter assumption is in contradiction with the possibility of the witnessing task introduced above.

In order for the witnessing task to be possible, a dynamical transformation $U_{MQ}$ must be allowed on the joint system $M\bigoplus Q$, that conserves the quantity $Z_M+Z_Q$ (as per condition (i)). 
\begin{figure}
    \centering
    \includegraphics[width=\columnwidth]{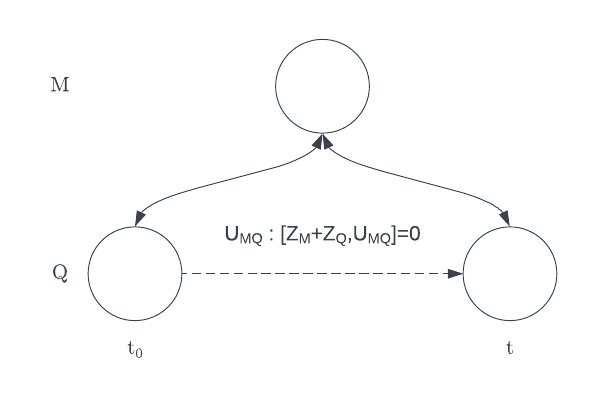}
    \caption{Pictorial representation of the temporal witness of non-classicality.}
    \label{fig:temporalwitnessscheme}
\end{figure}
This condition requires $\left[U_{MQ},Z_M+Z_Q\right]=0$. If $M$ just has one classical variable $Z_M$ and we consider the qubit algebra, then only Hamiltonians of the form $\alpha Z_Q+\beta Z_M+ \gamma Z_QZ_M$ are allowed, where $\alpha, \beta, \gamma$ are real-valued. Generalisations are possible to higher dimensions when we relax the assumptions of Q being a qubit and M being a bit. In these scenarios, the Hamiltonian describing the interactions between Q and M will change to include the generators of the appropriate algebra. However, the absence of two non-compatible degrees of freedom in M will still render this interaction incapable of making Q’s classical basis unsharp if it was sharp initially \cite{sharp}. Hence if $ M$ can realise the witnessing task, and the assumptions are satisfied, $M$ must have an extra, non-commuting variable -- thus being non-classical, see Fig.\ref{fig:temporalwitnessscheme}.

It is important to note here that the conservation law is enforced on the \textit{global} variable $Z_Q+Z_M$, rather than on the local variables $Z_Q$ and $Z_M$.
Furthermore, due to the additive form of the conserved quantity, the global system $Q\oplus M$ can be expanded to include additional subsystems, provided that their \textit{local} variables are included in the global conserved quantity. This makes the temporal witness of non-classicality more robust, as the model presented here can be easily extended to include an environment, provided that: 1) its local variables satisfy the global conservation law, and 2) there is a sufficiently high degree of control over its relevant degrees of freedom to exclude it from being responsible for $Q$'s coherent evolution.

In the exciton transfer scenario, the quantum controllable system $Q$ is a single \textit{photon} exciting the bio-polymer at $t_0$, while the system $M$ is the exciton in a molecular environment. The witnessing task is the quantum coherent evolution of the photon, mediated by the \textit{exciton} $M$ that is created on the bio-polymer when the photon $Q$ is absorbed. The conservation of the \textit{global energy} of the photon-exciton system and the molecular environment ensures that assumption (i) is satisfied. This scenario interestingly corresponds closely to experiments on exciton transfer in polydiacetylene, see e.g. \cite{dubin_macroscopic_2006}.

\section{A qubit model for the exciton transfer}

We shall now propose a quantum model to demonstrate the possibility of the witnessing task under energy conservation in a noisy environment. This will provide the best experimental conditions to implement the temporal witness of non-classicality in a real scenario.

We shall consider a bio-polymer whose $N$ monomers are described as a 1-d qubit chain.
Here ``monomer" refers to the smallest subunit of the complex system that can support an exciton alone. Experimentally, this can be initially selected according to the experimental setup, see e.g., \cite{dubin_macroscopic_2006}. In the Heisenberg picture, each monomer $M_j$ is described by its components $\{X_j,Y_j,Z_j\}$, $j=1,...,N$, satisfying the Pauli algebra, where ${Z_j}$ is the computational basis. We introduce the \textit{raising} and \textit{lowering} operators for each monomer, $\sigma^{\pm}_j=X_j \pm iY_j$, to describe them with the operators $\{\sigma^+_j, \sigma^-_j, Z_j\}$.

The chain's initial state is: 
\begin{equation}
    \rho_S=\ket{00...0}\bra{00...0}
\end{equation} as no excitation has been created yet. 

\textit{Creation of an exciton}.-- 
A photon $Q$ creates an exciton on the chain. It is initially localised on the \textit{first} monomer of the chain. Its quantum observables are ${X^p_j, Y^p_j, Z^p_j}$, $\alpha^\pm_j=X^p_j \pm iY^p_j$ and its initial state is:
\begin{equation}
    \rho_p=\ket{10...0}\bra{10...0}.
\end{equation}
Once the photon interacts with the first monomer, its degree of freedom is swapped with the degree of freedom of the monomer itself, so that:
\begin{equation}
    ^0\rho_S=Tr_{\rho_p}[\SWAP_{M_1,p} \rho_S \otimes \rho_p \SWAP^\dagger_{M_1,p}]=\ket{10...0}\bra{10...0}
    \label{eq:state0iter}\;.
\end{equation} Now the exciton is \textit{localised} on the first monomer of the chain. We focus on the single exciton regime, i.e., the probability of creating a second exciton in the chain is negligible \cite{dubin_macroscopic_2006}. Crucially, the conservation law is satisfied by the first step of the model: $\left[\SWAP_{M_1,p}, \sum_{j=1}^N{Z_j}+Z_p\right]=0$.

\textit{Exciton Dynamics and Environment}.-- 
We describe the exciton propagation with an \textit{XX-Hamiltonian} \cite{kay_perfect_2010}:
\begin{align}
    H_{XX}&=\frac{1}{2}\sum_{n=1}^NJ_n\left(X_nX_{n+1}+Y_nY_{n+1}\right)-\sum_{n=1}^N{B_nZ_n}
     & \\
    &=\frac{1}{4}\sum_{n=1}^NJ_n\left(\sigma^+_n\sigma^-_{n+1}+\sigma^-_n\sigma^+_{n+1}\right)-\sum_{n=1}^N{B_nZ_n} & \label{eq:XXHam}
\end{align} where $J_n, B_n \in \mathbb{R}$ $\forall\  n=1,...,N$. Notably, this model can capture different bio-polymers simply by changing $J_n$ and $B_n$. Thus it can be easily extended to other phenomena, like anisotropies in the bio-polymer, next-nearest neighbours coupling and multiple dimensions. Moreover, since $\left[H_{XX},\sum_{j=1}^N{Z_j}\right]=0$, this step of the model satisfies the conservation law too, as required by the witness.

To be realistic, one must also consider the environment, which induces decoherence. Here we shall model the environment as a thermal bath, using a \textit{quantum homogeniser} \cite{ziman_diluting_2002}. This is a reservoir of $R$ qubits that when suitably initialised can be used to prepare a system qubit in any quantum state,  to an arbitrarily high precision, improving as we increase $R$.
The \textit{only} unitary that can accomplish this task, called \textit{homogenisation}, is the \textit{partial swap} $\PSWAP=\cos{\eta}\ \mathbf{I}+i\sin{\eta}\ \SWAP$, where $\mathbf{I}$ is the identity operator on a two-qubit Hilbert space and $\eta$ is the interaction strength between the system and the reservoir. This makes the quantum homogeniser a \textit{universal quantum machine} \cite{ziman_diluting_2002}. Using this machine to describe the decoherence induced by a suitably prepared environment on the system, thus, doesn't affect the generality of the conclusions.

Initially, the qubits in the reservoir are all prepared in the same maximally mixed state, so that:
\begin{equation}
    ^0\xi=\bigotimes^{R}_{j=1} {^0\xi_j}=\left[\frac{1}{2}(\ket{0}\bra{0}+\ket{1}\bra{1})\right]^{\otimes R}.
    \label{eq:maxmixstate}
\end{equation} 
The decoherence is modelled by the interaction between the system and the reservoir, via the partial swap, and we shall call the interaction strength $\eta$ the \textit{decoherence strength}. With this initial state for the reservoir, a homogenisation of (at least) one of the monomers to the maximally mixed state would make that qubit effectively classical, thus not capable anymore of transmitting quantum coherence. Collision models are often applied to describe processes within a noisy environment, including decoherence. Crucially, they can offer a reliable estimation of the noise that is present in warm biological environments, {\cite{ziman_description_2005,ziman_all_2005,violaris_transforming_2021, civolani_engineering_2023}}. This unitary satisfies the conservation law because $\left[\PSWAP, \sum_{j=1}^N{Z_j}+\sum_{l=1}^R{Z_l}\right]=0$. 

\begin{figure}
    \centering
    \includegraphics[width=\columnwidth]{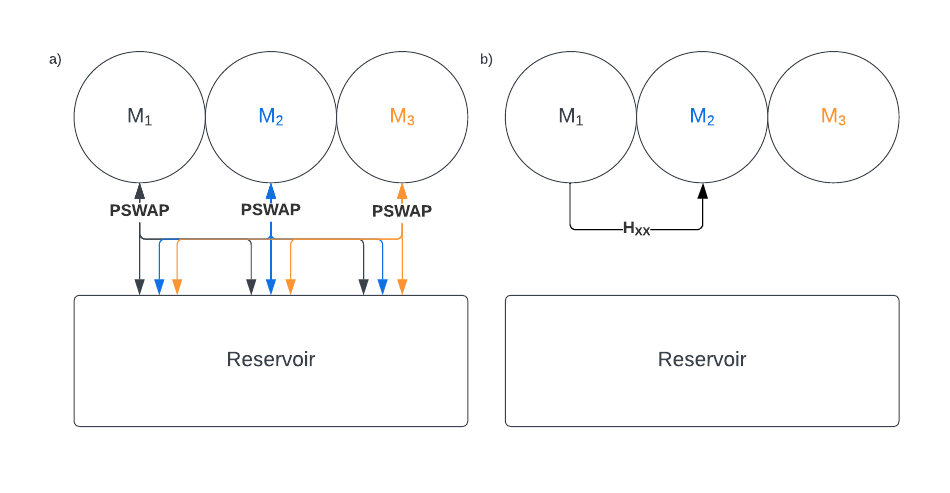}
    \caption{Schematic representation of the protocol: a) Decoherence phase; b) Transfer phase.}
    \label{fig:protocol}
\end{figure}
The protocol for the decohered exciton transfer works as follows. At each iteration of the protocol $k$, with $k=1,...,N$, \textit{each} monomer $M_j$, $j=1,...,N$, undergoes the homogenisation process with the $R$ qubits/phonons in the reservoir. This is the \textit{decoherence phase}, see Fig.\ref{fig:protocol}a). When the homogenisation has been performed on all the monomers, i.e., the decoherence round has ended, the monomer $M_k$ \textit{transfers} its quantum state to the monomer $M_{k+1}$. This is the \textit{transfer phase} of the protocol, see Fig.\ref{fig:protocol}b). After the transfer phase, the protocol can be repeated for the $(k+1)$-th iteration. 
This explains the notation: the upper script $0$ in $^0\rho_S$ and $^0\xi$ refers to the state of the system and the reservoir, respectively, \textit{before} the protocol begins, at $k=0$.

We shall introduce two different models for the reservoir in this scenario: (1) \textit{Markov Environment}: we re-initialise the quantum homogeniser in the maximally mixed state $^0\xi$ whenever a new monomer is involved in the decoherence phase of the protocol, at every iteration; (2) \textit{Non-Markov Environment}: the reservoir is \textit{never} re-initialised, neither when a new monomer enters the decoherence phase nor when a new iteration of the protocol begins.

\textit{Exciton recombination}.-- 
At the end of the $N^{th}$ iteration, we model recombination by swapping again the spatial degrees of freedom of the bio-polymer and that of the photon. The witnessing task is performed if the photon is \textit{quantum coherently delocalised} on every monomer of the chain:
\begin{align}
    \rho_p =&p_{0,0} \ket{00...00}\bra{00...00} + p_{1,0} \ket{10...00}\bra{00...00} + &  \nonumber \\ + & p_{0,1} \ket{00...00}\bra{10...00} +... + p_{N,N} \ket{00...01}\bra{00...01}; &\;
    \label{eq:finalstatephoton}
\end{align} This dynamics provides the coherent evolution required by the temporal witness of non-classicality.

To determine experimentally whether the witnessing task is achieved, one can measure the position of the photon with an interference experiment: observing interference fringes and assuming the conservation law of the additive quantity $\sum_{j=1}^N{Z_j} + \sum_{l=1}^R{Z_l} + Z_p$, the witness allows us to conclude that the exciton, and hence the bio-polymer, is non-classical. We stress here that the conservation law is enforced on the \textit{whole} system made by the $N$ monomers, the $R$ reservoir qubits and the photon. This means that one can consider a change in the $Z$ component of one subsystem if it is balanced by an equal and opposite change in the $Z$ component of the other subsystems.

\subsection{Markovian vs Non-Markovian Environment}
Here we derive the final state of the photon in the framework of both a \textit{Markovian} and a \textit{Non-Markovian} environment. The former is equivalent to having a different reservoir per monomer at each iteration of the protocol, the latter to a reservoir evolving with the bio-polymer. This is the Non-Markovian feature of the model (different from other models, e.g. in \cite{saha_quantum_2024}): the state of each reservoir qubit has ``memory", and it cannot be re-initialised at every iteration.
We assume for simplicity that the system is an $N=3$ monomer chain and the reservoir is made of $R=3$ qubits. Every reservoir is initialised in the maximally mixed state in Eq.\ref{eq:maxmixstate}, while the bio-polymer is initially in the state in Eq.\ref{eq:state0iter}.
The final state of the photon after the exciton recombination in the Markovian scenario is \cite{mark}:
\begin{align}
    ^3\rho_p=&\left[{^3\rho_{p_1}}\otimes{^3\rho_{p_2}} + itJ_1\cos^{18}{\eta}\left(\alpha^+_1\alpha^-_2 - \alpha^-_1\alpha^+_2\right) \nonumber \right. & \\ 
    & \left. -t^2B_2 \left(\alpha^+_1\alpha^-_2 + \alpha^-_1\alpha^+_2\right)\right]\otimes{^3\rho_{p_3}} \nonumber & \\
    & -t^2J_1J_2\cos^{18}{\eta}\left(Z_2+\cos^{12}{\eta}\right)\left(\alpha^+_1\alpha^-_3 + \alpha^-_1\alpha^+_3\right),
    \label{eq:finalstateMarkov}
\end{align} while in the Non-Markovian scenario is \cite{non-mark}:
\begin{align}
    &^3\rho_p={^3\rho_{p_1}}\otimes{^3\rho_{p_2}}\otimes{^3\rho_{p_3}} \nonumber & \\
    & + itJ_1\left[F(\eta)+2itB_2G(\eta)\right]\left(\alpha^+_1\alpha^-_2 - \alpha^-_1\alpha^+_2\right)\otimes{^3\rho_{p_3}} \nonumber  & \\ 
    &  +itJ_1\left[G(\eta)+2itB_2F(\eta)\right] \left(\alpha^+_1\alpha^-_2 + \alpha^-_1\alpha^+_2\right)\otimes{^3\rho_{p_3}} \nonumber & \\
    & -\frac{1}{2}t^2J_1J_2\left(Z_2+\cos^{12}{\eta}+s(\eta)\right)\left[F(\eta)\left(\alpha^+_1\alpha^-_3 + \alpha^-_1\alpha^+_3\right) \right. \nonumber & \\ 
    & \left. +G(\eta)\left(\alpha^+_1\alpha^-_3 - \alpha^-_1\alpha^+_3\right)\right]
    \label{eq:finalstateNonMarkov}
\end{align}
where ${^3\rho_{p_j}}$ is the density matrix of the photon localised on the monomer $M_j$ at the $k=3$ iteration of the protocol, $F(\eta), G(\eta)$ and $s(\eta)$ are defined in Supplementary Eq.76, Supplementary Eq.77 and Supplementary Eq.79, respectively.

\section{Discussion}
We shall now compare the two states for the Markov (Eq.\ref{eq:finalstateMarkov}) and Non-Markov (Eq.\ref{eq:finalstateNonMarkov}) environments.

 In the weak coupling limit, in both scenarios, the exciton mediates a coherent delocalisation of the photon over the bio-polymer. Hence the witnessing task is successfully achieved, even in the presence of decoherence. Using the temporal witness of non-classicality, one can conclude that the exciton mediating the photon coherent delocalisation is non-classical, and so the process governing its transfer along the bio-polymer must be non-classical itself. 
 This is the key result of the work, which opens the possibility of applying the temporal witness of non-classicality introduced here to the field of quantum biology, providing a novel tool to discriminate conclusively the role of quantum effects in biological processes.
 
The model explains why, despite decoherence, a coherent delocalisation of the exciton over the chain is still possible: the interactions between the monomers mitigate the decoherence, preventing $^k\rho_S$ from becoming an eigenstate of $H_{XX}$ in Eq.\ref{eq:XXHam}.
\begin{figure}
    \centering
    \includegraphics[width=\columnwidth]{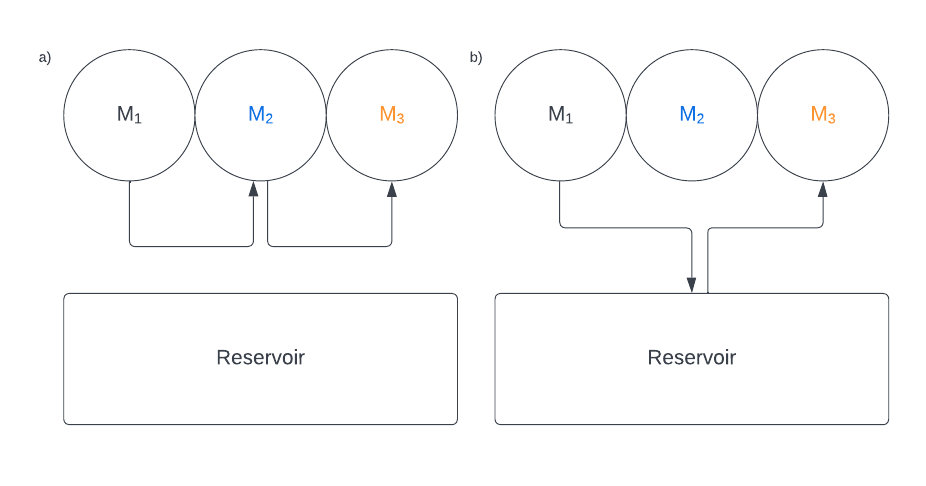}
    \caption{Schematic view of the coherent hopping mechanisms: a) via intermediate monomers; b) via reservoir.}
    \label{fig:mechanisms}
\end{figure}
Interestingly, at every time step, the exciton can face three different scenarios: (1) It can remain localised on the monomer preceding that involved in the transfer phase, as suggested by the coefficients $B_j$;
(2) It can (coherently) hop on to the next monomer. This can occur via two mechanisms: (2.1) Via the intermediate monomers, see Fig.\ref{fig:mechanisms}a). This is mathematically expressed by the terms in Eq.\ref{eq:finalstateMarkov} and Eq.\ref{eq:finalstateNonMarkov} that are proportional to $Z_j$: since $Z_j$ is conserved throughout the process, the exciton moves across the monomer $M_j$ leaving it globally unaltered;
(2.2) Via the reservoir, see Fig.\ref{fig:mechanisms}b).
One can see this mechanism in Eq.\ref{eq:finalstateMarkov} and in Eq.\ref{eq:finalstateNonMarkov} in the terms proportional to $\cos^{12}{\eta}$ and $\cos^{12}{\eta}+s(\eta)$, respectively. Recalling that the interaction between the polymer and the reservoir is described as a \PSWAP, for the transfer to occur in a Markovian scenario, the \SWAP\ between the reservoir qubit and the monomer must \textit{not} be performed at any step as this would put the monomer in a maximally mixed state, blocking the coherent evolution of the exciton on the polymer. This occurs with probability $\cos^2{\eta}$ at every protocol's iteration. In a fully non-Markovian scenario, instead, the environment can also exchange information with the chain, leading to active involvement in the coherent evolution of the exciton, which is described by the additional $s(\eta)$.
We notice that the preceding monomers are not involved here: the environment is solely responsible for the quantum coherent delocalisation of the exciton. 

The difference between a Markov and a Non-Markov environment, therefore, lies in the amount of coherence maintained by the qubits. Such coherence (i) creates, at every iteration, a second ``hopping term" between the two monomers involved in the transfer phase, shifted by a phase factor $e^{i\pi}$; (ii) reshapes the probabilities for the coherent delocalisation of the exciton over the bio-polymer.

Consider now the coefficients of the terms describing the aforementioned mechanisms. 

\begin{figure}
    \centering
    \includegraphics[width=0.8\columnwidth]{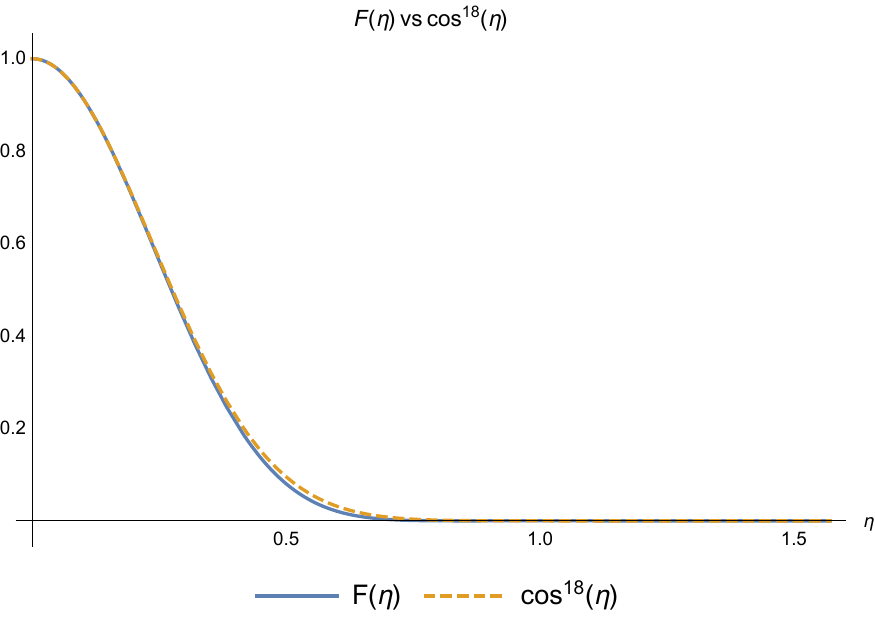}
    \caption{Coefficients $F(\eta)$ (blue) and $\cos^{18}{\eta}$ (orange) at the end of $k=2$ iteration}
    \label{fig:FvsCos}
\end{figure}
\begin{figure}
    \centering
    \includegraphics[width=0.8\columnwidth]{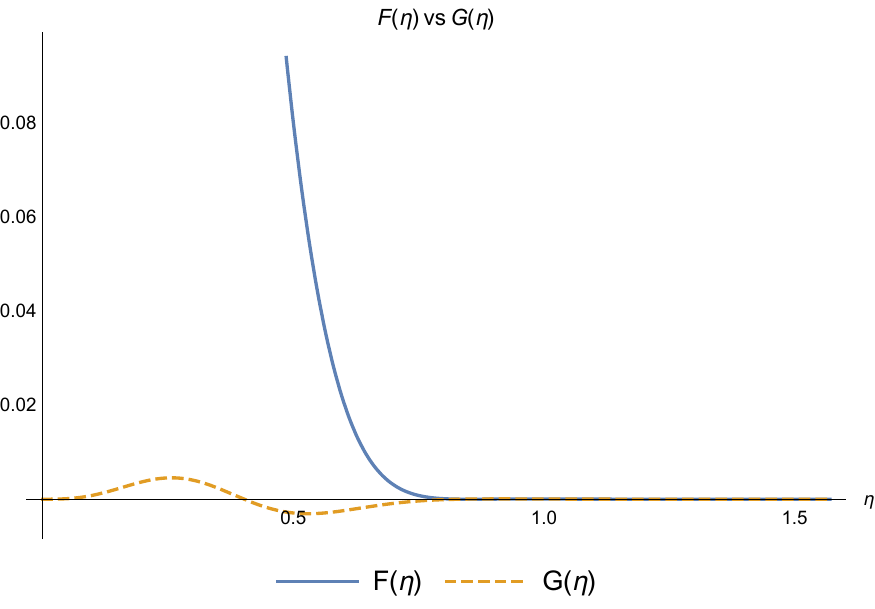}
    \caption{Coefficients $F(\eta)$ (blue) and $G(\eta)$ (orange) at the end of $k=2$ iteration.}
    \label{fig:FvsG}
\end{figure}

Concerning the coefficients describing the coherent delocalisation of the exciton \textit{via the chain} (mechanism (2.1)), Fig.\ref{fig:FvsCos} and Fig.\ref{fig:FvsG} show that in the weak coupling regime (i) $F(\eta)$ and the coefficient in the Markov scenario are equal, and this holds true at every step of the protocol and (ii) $G(\eta)$ appears to be very small, almost negligible, thus its effect is \textit{practically} negligible \cite{table}: small values of $\eta$ make the interaction with the environment very unlikely, therefore it cannot mediate the exciton transport along the bio-polymer. This elucidates why the coefficient $F(\eta)$ has the same value as its Markovian counterpart at every step of the protocol. 

Moreover, for some values of $\eta$, both $F(\eta)$ and $G(\eta)$ are vanishing: the hopping via chain is \textit{prevented} by the strong coupling with the environment. The effect of decoherence is so strong that the monomers lose coherence as soon as they interact with the environment. They effectively become equivalent to classical bits and not capable of quantum-coherently transmitting the exciton anymore.

\begin{figure}
    \centering
    \includegraphics[width=0.8\columnwidth]{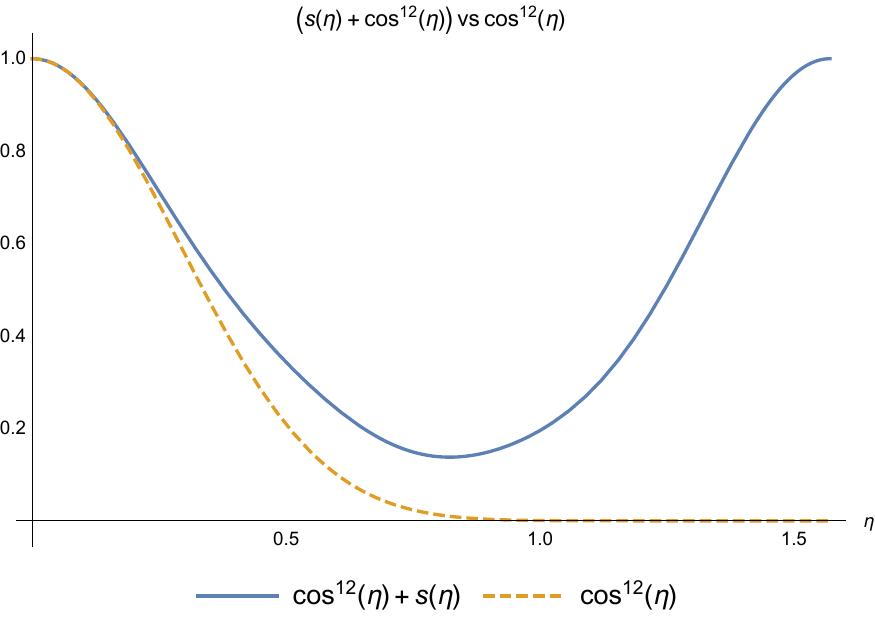}
    \caption{Coefficients $s(\eta)+\cos^{12}(\eta)$ (blue) and $\cos^{12}(\eta)$ (orange) at the end of $k=2$ iteration.}
    \label{fig:a_vs_cos}
\end{figure}

Considering the role played by the non-Markovian environment in (2.2), i.e., the hopping mechanism \textit{via the reservoir}, the same argument as before applies, focusing on the coefficients $s(\eta)$ and $\cos^{12}{\eta}$. They have almost the same numerical values as in the \textit{weak coupling} regime, as one can see in Fig.\ref{fig:a_vs_cos}.
As soon as the coupling with the environment increases, entering the \textit{strong coupling} regime, the hopping via reservoir becomes impossible with a Markovian environment, but it is still possible with a Non-Markov environment. Informally, one can say that the coherence ``stored" by a Non-Markov environment is then used by the bio-polymer for the quantum coherent delocalisation of the exciton.
\begin{figure}
    \centering
    \includegraphics[width=0.8\columnwidth]{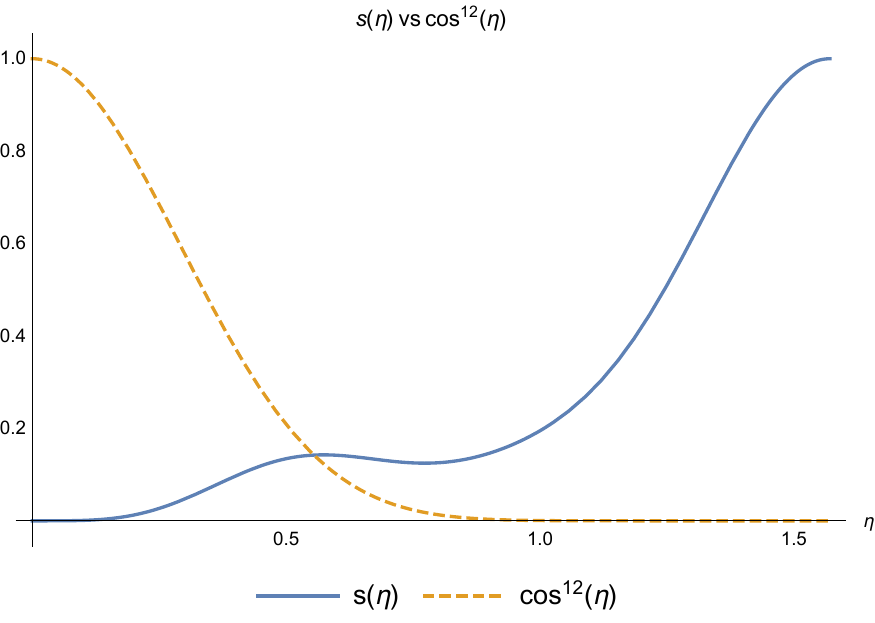}
    \caption{Coefficients $s(\eta)$ (blue) and $\cos^{12}(\eta)$ (orange) at the end of $k=2$ iteration.}
    \label{fig:s_vs_cos}
\end{figure}
This is clarified by Fig.\ref{fig:s_vs_cos}: entering the strong coupling regime, the coefficient $s(\eta)$ describing the coherence exchange between system and environment becomes larger than the $\cos^{12}{\eta}$, which instead describes a passive role of the environment in the exciton transport on the chain. Hence the non-Markovian environment becomes \textit{crucial} for the energy transfer in the strong coupling regime, as no other hopping mechanism is allowed. This may explain why bio-polymer vibrations may be relevant in exciton-mediated transport phenomena, such as light-harvesting \cite{butkus_vibrational_2012, tempelaar_vibrational_2014, reppert_quantumness_2018, duan_origin_2015, halpin_two-dimensional_2014}.

\section{Conclusions}

We have proposed a novel experimental scheme to witness non-classicality in the exciton transfer on a bio-polymer, using a photonic quantum probe. This witness is subtly different from other ones, e.g. \cite{wilde_could_2009,li_witnessing_2012, knee_subtleties_2018}, as it investigates an \textit{intrinsic}, more \textit{general}, property of the biological system itself, rather than of the specific process. This witness has the appeal of being very broadly applicable and of relying only on a few general assumptions: energy conservation and the formalism of quantum theory.
One in fact could relax the latter and recast this witness of non-classicality in the information-theoretic framework known as \textit{constructor theory of information} \cite{deutsch_constructor_2015}. This dynamics-independent approach would achieve full generality in our argument, giving interesting insights regarding possible \textit{physical} reasons why biological systems must be non-classical \cite{marletto_constructor_2015}. We shall explore this in a forthcoming paper.

We have also provided a qubit model to discuss the possibility of the witnessing task, namely the quantum coherent evolution of the photon probe, with an analysis of decoherence, in the Markovian and non-Markovian regimes. The model shows that the witnessing task is always possible, even in the presence of decoherence, and it does so in a system and scale-independent way. The generality of the model makes it suitable to describe several existing quantum biology results, \cite{engel_evidence_2007,strumpfer_effect_2011,renger_normal_2012}. This shows that the application of the temporal witness of non-classicality \cite{di_pietra_temporal_2023} is meaningful in the field of quantum biology and it could shed new light on the importance of quantum effects in biological systems. Future investigations on the possibility of designing an experiment with a real biological system can benefit from the results discussed in this work, using this model as a starting point to be enriched with system-specific features.
For example, the structure of the environment presented in this work is similar to the one considered in \cite{olbrich_theory_2011}, where molecular dynamics simulations are employed to study the coupling between the protein environment and the vertical excitation energies in a real biological system. Enriched with the results discussed in this paper, the same simulation could be used to explore the applications of the temporal witness of non-classicality to a more realistic system.
 
The role of a Non-Markovian environment in the witnessing task becomes essential in the strong coupling regime: the only way for the witnessing task to be possible when $\eta=\frac{\pi}{2}\approx1.571 $ is to rely on the interaction of the bio-polymer with its own \textit{Non-Markovian} environment. This is because the reservoir has memory of the occurred process and maintains coherence throughout it. Instead, in a weak coupling scenario, the witnessing task is possible \textit{independently of the Markovianity} of the reservoir. This conclusion is intuitively pleasing since in the strong coupling regime the separation between the system and environment becomes nonphysical and it is more appropriate to think of the environment as being part of the system. 
 
The generality of our witnessing scheme makes it an ideal candidate for designing new experiments in quantum biology, to pin down conclusively the role of quantum effects in exciton transfer and other energy transfer processes. We leave the development of this research to future work, to enrich this idealised model with system-specific dynamical features to make it capable of informing the \textit{feasibility} of the witnessing task in a real system, not only its possibility. 
 

{\bf Data Availability} \;\; All data generated or analysed during this study are included in this published article and its Supplementary Information file.

{\bf Acknowledgements} \;\; We thank Simone Rijavec, Maria Violaris, Mattheus Burkhard, Antonio Pantelias Garcés, Virginia Tsiouri and Tristan Farrow for sharp comments and fruitful discussions on this manuscript. This research was made possible through the generous support of the Gordon and Betty Moore Foundation. G.D.P. thanks the Clarendon Fund and the Oxford-Thatcher Graduate Scholarship for supporting this research.

\bibliographystyle{apsrev4-2}
\bibliography{bibpolymer}

\begin{widetext}
\input{supplementary}
\end{widetext}
\end{document}

%% file: supplementary.tex
\section{Markovian Environment: analytical derivation of Eq.\ref{eq:finalstateMarkov}}

In this Appendix, we will analytically derive Eq.\ref{eq:finalstateMarkov} and the general state for $N$ monomers and $R$ reservoir qubits, going iteration by iteration in the protocol and providing the final state of the emitted photon after the recombination of the exciton on the polymer. The setup is the very same discussed in Section \textit{Markovian vs Non-Markovian Environment}.

\textit{First iteration $k=1$.}--
We start with the polymer being in the state described by the density matrix in Eq. \ref{eq:state0iter}. At this stage, the state of the system is separable, so that we can write:
\begin{equation}
    ^0\rho_S={^0\rho_{S_1}}\otimes{^0\rho_{S_2}}\otimes...\otimes{^0\rho_{S_N}}
\end{equation} being $^0\rho_{S_j}$, $j=1,...,N$, the reduced density matrix of the monomer $M_j$ \textit{before} the beginning of the homogenisation process in the $k=1$ iteration. 
It is convenient to write the single monomer density matrices using the Bloch vector representation:
\begin{equation}
    ^0\rho_{S_1}=\frac{1}{2}\left(\mathbf{I}+{^0\Vec{s}_1}\cdot\Vec{\sigma}\right)=\frac{1}{2}\left(\mathbf{I}-Z_1\right)
\end{equation} where $\sigma$ is the vector of Pauli operators and the initial condition for the first monomer of the chain is $^0\Vec{s}_1=-\hat{z}$ since it is excited by the photon, so it is in the state $\ket{1}$. Similarly:
\begin{equation}
    ^0\rho_{S_j}=\frac{1}{2}\left(\mathbf{I}+{^0\Vec{s}_j}\cdot\Vec{\sigma}\right)=\frac{1}{2}\left(\mathbf{I}+Z_j\right),\ \ \ j=2,...,N
\end{equation} where the initial conditions are $^0\Vec{s}_j=+\hat{z}$ because all the remaining monomers are \textit{not} excited by the photon at the beginning. Recall in fact that our photon is initially spatially localised on the first monomer.
Moving now to the reservoir qubits, we can adopt the Bloch vector representation for them too:
\begin{equation}
    ^0\xi_{j}=\frac{1}{2}\left(\mathbf{I}+{^0\Vec{r}_j}\cdot\Vec{\sigma}\right)=\frac{1}{2}\left(\mathbf{I}\right),\ \ \ j=1,...,R
\end{equation} where $^0\Vec{r}_j=0$ because we initialise all the qubits in the reservoir in the maximally mixed state.

We can now start with the decoherence phase of the $k=1$ iteration of the protocol, i.e., the first round of homogenisation for the polymer. We start with the first, excited, monomer. It interacts with the first qubit in the reservoir:
\begin{equation}
    ^1\rho^{(1)}_{S_1}=Tr_{{^0\xi_1}}\left[\PSWAP_{{^0\rho_{S_1}},{^0\xi_1}}  {\left({^0\rho_{S_1}}\otimes{^0\xi_1}\right)} \PSWAP^\dagger_{{^0\rho_{S_1}},{^0\xi_1}}\right]= \frac{1}{2}\left(\mathbf{I}+ {^1\Vec{s}^{(1)}_1}\cdot\Vec{\sigma}\right)= \frac{1}{2}\left(\mathbf{I}-\cos^2{\eta}\ Z_1\right) 
\end{equation} where the Bloch vector ${^1\Vec{s}^{(1)}_1}$ evolves as:
\begin{equation}
    ^1\Vec{s}^{(1)}_1=\cos^2{\eta}\ {^0\Vec{s}_1} +\sin^2{\eta}\ {^0\Vec{r}_1} -\cos{\eta}\sin{\eta}\left({^0\Vec{r}_1} \times {^0\Vec{s}_1}\right)=-\cos^2{\eta}\ \hat{z}. 
\end{equation} This means that the decoherence process, at this stage of the protocol, shrinks the length of the monomer's Bloch vector of a factor $\cos^2{\eta}$ for every qubit in the reservoir.
We can easily generalise this process to the $R^{th}$ interaction with the reservoir, to have:
\begin{equation}
    ^1\rho^{(R)}_{S_1}= \frac{1}{2}\left(\mathbf{I}+ {^1\Vec{s}^{(R)}_1}\cdot\Vec{\sigma}\right)= \frac{1}{2}\left(\mathbf{I}-\cos^{2R}{\eta}\ Z_1\right)
\end{equation} where:
\begin{equation}
    ^1\Vec{s}^{(R)}_1=\cos^2{\eta}\ {^1\Vec{s}^{(R-1)}_1} +\sin^2{\eta}\ {^0\Vec{r}_R}-\cos{\eta}\sin{\eta}\left({^0\Vec{r}_R} \times {^0\Vec{s}^{(R-1)}_1}\right)=\cos^2{\eta}(-\cos^{2(R-1)}{\eta}) \hat{z}=-\cos^{2R}{\eta}\ \hat{z}.
    \label{eq:inicond1k2}
\end{equation} This concludes the decoherence phase for the first monomer.

We move now to the second monomer. Since we are in the Markov environment scenario, we can initialise again the reservoir to a collection of $R$ qubits prepared in the maximally mixed state, or equivalently we can say that the second monomer interacts with a different environment, equally built and prepared. This allows us to use the very same results we got for the decoherence phase of the first monomer, with the only difference that the initial condition for the second monomer $M_2$ is given by $^0\Vec{s}_2=+\hat{z}$ since it is not excited by the photon. We will thus have:
\begin{equation}
    ^1\rho^{(R)}_{S_2}= \frac{1}{2}\left(\mathbf{I}+ {^1\Vec{s}^{(R)}_2}\cdot\Vec{\sigma}\right)= \frac{1}{2}\left(\mathbf{I}+\cos^{2R}{\eta}\ Z_2\right) 
\end{equation} where:
\begin{equation}
    ^1\Vec{s}^{(R)}_2=\cos^2{\eta}\ {^1\Vec{s}^{(R-1)}_2} +\sin^2{\eta}\ {^0\Vec{r}_R} -\cos{\eta}\sin{\eta}\left({^0\Vec{r}_R} \times {^0\Vec{s}^{(R-1)}_2}\right)=\cos^2{\eta}(+\cos^{2(R-1)}{\eta}) \hat{z}=+\cos^{2R}{\eta}\ \hat{z}. 
    \label{eq:inicond2k2}
\end{equation} 

The same reasoning can be applied to all the remaining monomers in the chain, so that at the end of the decoherence phase for the $k=1$ iteration of the protocol our monomers will be in the states:
\begin{equation}
\begin{cases}
    ^1\rho^{(R)}_{S_1}=\frac{1}{2}\left(\mathbf{I}-\cos^{2R}{\eta}\ Z_1\right) \\
    ^1\rho^{(R)}_{S_j}=\frac{1}{2}\left(\mathbf{I}+\cos^{2R}{\eta}\ Z_j\right)\ \ \ \ \ j=2,...,N    
\end{cases}
\end{equation} and the state of the polymer:
\begin{equation}
    ^1\rho_S={^1\rho^{(R)}_{S_1}}\otimes {^1\rho^{(R)}_{S_2}}\otimes ... \otimes {^1\rho^{(R)}_{S_N}}.
\end{equation}

The first iteration of the protocol ends with the \textit{transfer phase} involving the monomers $M_1$ and $M_2$. This is described by the unitary ($\hbar=1$ from now on):
\begin{equation}
    U=e^{-iH_{XX}t}
    \label{eq:unitary}
\end{equation}
where $H_{XX}$ is the XX Hamiltonian in Eq.\ref{eq:XXHam}. We find:
\begin{equation}
    ^1\rho_{S_{1,2}}=U\left( {{^1\rho^{(R)}_{S_1}} \otimes {^1\rho^{(R)}_{S_2}}}\right) U^\dagger
\end{equation} which, at first order in $t$, gives:
\begin{equation}
    ^1\rho_{S_{1,2}}={^1\rho^{(R)}_{S_1}} \otimes {^1\rho^{(R)}_{S_2}} + \frac{1}{4}itJ_1\cos^{2R}{\eta}\left(\sigma^+_1\sigma^-_2 - \sigma^-_1\sigma^+_2\right).
    \label{eq:rho12}
\end{equation} 
All in all, the state of the polymer after the $k=1$ iteration of the protocol, which will in turn be the initial state for the $k=2$ iteration, is:
\begin{equation}
    ^2\rho_S={^1\rho_{S_{1,2}}}\otimes{^1\rho^{(R)}_{S_j}}^{\otimes (N-2)}\ \ \ j=3,...,N.
    \label{eq:state1iter}
\end{equation}

\textit{Second iteration $k=2$}.--
Since we are dealing with the Markov environment, before beginning with the decoherence phase we must initialise again the reservoir in the state $^0\xi$ of Eq.\ref{eq:maxmixstate}. The state of the polymer is the one in Eq.\ref{eq:state1iter} instead, where the monomers $M_1$ and $M_2$ are entangled. In this case, the homogenisation process will be performed \textit{locally} on both $M_1$ and $M_2$. The entangled state $^1\rho_{S_{1,2}}$ in Eq.\ref{eq:rho12} can be divided into a separable and a non-separable part, which we can discuss separately.

The separable part of Eq.\ref{eq:rho12} will evolve as described in the $k=1$ iteration of the protocol, with the only difference that now the initial conditions for the two monomers $M_1$ and $M_2$ are given by $^1\Vec{s}^{(R)}_1$ in Eq.\ref{eq:inicond1k2} and $^1\Vec{s}^{(R)}_2$ in Eq.\ref{eq:inicond2k2}. After $R$ interactions with the reservoir, we will have;
    \begin{equation}
        \begin{cases}
            ^2\rho^{(R)}_{S_1}=\frac{1}{2}\left(\mathbf{I}-\cos^{(4R)}{\eta} Z_1\right) \\
            ^2\rho^{(R)}_{S_2}=\frac{1}{2}\left(\mathbf{I}+\cos^{(4R)}{\eta} Z_2\right)
        \end{cases}
    \end{equation}

The \textit{non}-separable part is the most interesting to explore. We shall call it $^1\rho^{non-sep}_{S_{1,2}}$. Let us start with the first monomer being involved in the decoherence phase. Interacting with the first qubit of the reservoir, we find:
\begin{equation}
    ^1\rho^{non-sep, (1)}_{S_{1,2}}=\frac{1}{2}itJ_1\cos^{2R}{\eta}\cos^2{\eta}\left(\sigma^+_1\sigma^-_2 - \sigma^-_1\sigma^+_2\right)
    \label{eq:nonsepMark}
\end{equation} so that, after the interaction with the remaining $R-1$ qubits in the reservoir, $^1\rho^{non-sep}_{S_{1,2}}$ will be:
\begin{equation}
    ^1\rho^{non-sep, (R)}_{S_{1,2}}=2^{(R-2)}itJ_1\cos^{4R}{\eta}\left(\sigma^+_1\sigma^-_2 - \sigma^-_1\sigma^+_2\right).
\end{equation} The same reasoning can be applied when the second monomer $M_2$ undergoes the decoherence phase. After the interaction with the $R-th$ qubit in the reservoir, the final state of the non-separable part of $^1\rho_{S_{1,2}}$ is:
\begin{equation}
    ^1\rho^{non-sep, (2R)}_{S_{1,2}}=2^{(2R-2)}itJ_1\cos^{6R}{\eta}\left(\sigma^+_1\sigma^-_2 - \sigma^-_1\sigma^+_2\right).
\end{equation}

All in all, the entangled state in Eq.\ref{eq:rho12} after the decoherence phase of the $k=2$ iteration becomes (re-scaling the coefficient properly to remove the constant numeric factors):
\begin{equation}
    ^2\rho^{(R)}_{S_{1,2}}={^2\rho^{(R)}_{S_1}}\otimes{^2\rho^{(R)}_{S_2}} +itJ_1\cos^{6R}{\eta}\left(\sigma^+_1\sigma^-_2 - \sigma^-_1\sigma^+_2\right),
\end{equation} while the remaining $N-2$ monomers, not being involved in the quantum state transfer, evolve as described in the previous iteration:
\begin{equation}
    ^2\rho^{(R)}_{S_j}=\frac{1}{2}\left(\mathbf{I}+\cos^{4R}{\eta}Z_j\right)\ \ \ \ j=3,...,N.
\end{equation}

We move now to the transfer phase of the $k=2$ iteration, involving the monomers $M_2$ and $M_3$. Applying again the unitary in Eq.\ref{eq:unitary} to first order in $t$, we find:
\begin{align}
    ^2\rho_{S_{1,2,3}}=U\left( {{^2\rho^{(R)}_{S_{1,2}}} \otimes {^2\rho^{(R)}_{S_3}}}\right)U^\dagger=&\left[{^2\rho^{(R)}_{S_1}}\otimes{^2\rho^{(R)}_{S_2}} + itJ_1\cos^{6R}{\eta}\left(\sigma^+_1\sigma^-_2 - \sigma^-_1\sigma^+_2\right) - t^2B_2\left(\sigma^+_1\sigma^-_2 + \sigma^-_1\sigma^+_2\right) \right]\otimes {^2\rho^{(R)}_{S_3}} & \nonumber \\
    & - t^2 J_1J_2\cos^{6R}{\eta}\left(Z_2+\cos^{4R}{\eta}\right)\left(\sigma^+_1\sigma^-_3 + \sigma^-_1\sigma^+_3\right). & \label{eq:rho123}
\end{align}
We can now summarise the outcome of the $k=2$ iteration of the protocol, which in turn is the initial state of the iteration $k=3$. The state of the polymer will be:
\begin{equation}
    ^3\rho_S={^2\rho_{S_{1,2,3}}}\otimes{^2\rho^{(R)}_{S_j}}^{\otimes (N-3)}\ \ \ j=4,...,N.
    \label{eq:state2iter}
\end{equation}

\textit{$N-th$ iteration $k=N$ and recombination}.--
Once we have understood how our decoherence map works for the decoherence phase and what mechanisms are involved in the transfer phase, we can easily generalise the protocol to the very last step: the exciton arrives at the last but one monomer in the chain and, after the last decoherence phase, is transmitted to the monomer $M_N$; it is at this stage that the exciton will finally recombine on the polymer, emitting again a single photon. The final state of the photon will thus be:
\begin{align}
    &^N\rho_p=\left[{^N\rho_{p_1}}\otimes{^N\rho_{p_2}}+ itJ_1\cos^{[2+4(N-2)]R}{\eta}\left(\alpha^+_1\alpha^-_2 - \alpha^-_1\alpha^+_2\right)-t^2B_2\cos^{4(N-3)R}{\eta}\left(\alpha^+_1\alpha^-_2 + \alpha^-_1\alpha^+_2\right)\right]{^N\rho_{p_3}}\otimes...\otimes{{^N\rho_{p_N}}} \nonumber & \\
    & - t^2 J_1 J_2\cos^{[2+4(N-2)]R}{\eta}\left(\cos^{(N-3)2R}{\eta}Z_2+\cos^{4R}{\eta}\right)\left[\left(\alpha^+_1\alpha^-_3 + \alpha^-_1\alpha^+_3\right) + itB_3 \left(\alpha^+_1\alpha^-_3 - \alpha^-_1\alpha^+_3\right)\right]{^N\rho_{p_4}}\otimes...\otimes{{^N\rho_{p_N}}} \nonumber & \\ 
    & +...+(i)^{2N-3}t^{N-2}J_1J_2...J_{N-1}\cos^{[2+4(N-2)]R}{\eta}\left(\cos^{(N-3)2R}{\eta}Z_2+\cos^{4R}{\eta}\right)...\left(Z_N+\cos^{2(N-1)R}{\eta}\right) \cdot \nonumber & \\
    & \cdot \left[\left(\alpha^+_1\alpha^-_N + (-1)^{2N-3} \alpha^-_1\alpha^+_N\right) + itB_N \left(\alpha^+_1\alpha^-_N + (-1)^{2N-2} \alpha^-_1\alpha^+_N\right)\right].
    \label{eq:finalstateMarkovgeneral}
\end{align} where ${^N\rho_{p_j}}$ is the density matrix of the photon localised over the monomer $M_j$ at the $k=N$ iteration of the protocol.

\section{Non-Markovian Environment: analytical derivation of Eq.\ref{eq:finalstateNonMarkov}}

Here we will detail the procedure leading to Eq.\ref{eq:finalstateNonMarkov} and provide some useful generalisation to compute the needed coefficients for larger systems. 
The setup is the very same as discussed in Section \textit{Markovian vs Non-Markovian Environment}.

\textit{First iteration $k=1$}.--
We start again with the polymer being in the state described by the density matrix in Eq. \ref{eq:state0iter}:
\begin{equation}
    ^0\rho_S={^0\rho_{S_1}}\otimes{^0\rho_{S_2}}\otimes{^0\rho_{S_3}}
\end{equation} being $^0\rho_{S_j}$, $j=1,...,3$, the reduced density matrix of the monomer $M_j$ \textit{before} the beginning of the homogenisation process in the $k=1$ iteration. 
We write the single monomer density matrices using the Bloch vector representation:
\begin{equation}
    ^0\rho_{S_1}=\frac{1}{2}\left(\mathbf{I}+{^0\Vec{s}_1}\cdot\Vec{\sigma}\right)=\frac{1}{2}\left(\mathbf{I}-Z_1\right)
\end{equation} where $\sigma$ is the vector of Pauli operators and the initial condition for the first monomer of the chain is $^0\Vec{s}_1=-\hat{z}$ since it is excited by the photon, so it is in the state $\ket{1}$. Similarly:
\begin{equation}
    ^0\rho_{S_j}=\frac{1}{2}\left(\mathbf{I}+{^0\Vec{s}_j}\cdot\Vec{\sigma}\right)=\frac{1}{2}\left(\mathbf{I}+Z_j\right),\ \ \ j=2,3
\end{equation} where the initial conditions are $^0\Vec{s}_j=+\hat{z}$ because all the remaining monomers are \textit{not} excited by the photon at the beginning. Recall in fact that our photon is initially spatially localised on the first monomer.
Moving now to the reservoir qubits, we can adopt for them too the Bloch vector representation:
\begin{equation}
    ^0\xi_{j}=\frac{1}{2}\left(\mathbf{I}+{^0\Vec{r}_j}\cdot\Vec{\sigma}\right)=\frac{1}{2}\left(\mathbf{I}\right),\ \ \ j=1,...,3
\end{equation} where $^0\Vec{r}_j=0$ because we initialise all the qubits in the reservoir in the maximally mixed state.

We can now start with the \textit{decoherence phase} of the $k=1$ iteration of the protocol, i.e., the first round of homogenisation for the polymer. We start with the first, excited, monomer. It interacts with the first qubit in the reservoir:
\begin{equation}
    ^1\rho^{(1)}_{S_1}=Tr_{{^0\xi_1}}\left[\PSWAP_{{^0\rho_{S_1}},{^0\xi_1}}  {\left({^0\rho_{S_1}}\otimes{^0\xi_1}\right)} \PSWAP^\dagger_{{^0\rho_{S_1}},{^0\xi_1}}\right]= \frac{1}{2}\left(\mathbf{I}+ {^1\Vec{s}^{(1)}_1}\cdot\Vec{\sigma}\right)= \frac{1}{2}\left(\mathbf{I}-\cos^2{\eta}\ Z_1\right) 
\end{equation} where the Bloch vector ${^1\Vec{s}^{(1)}_1}$ evolves as:
\begin{equation}
    ^1\Vec{s}^{(1)}_1=\cos^2{\eta}\ {^0\Vec{s}_1} +\sin^2{\eta}\ {^0\Vec{r}_1} -\cos{\eta}\sin{\eta}\left({^0\Vec{r}_1} \times {^0\Vec{s}_1}\right)=-\cos^2{\eta}\ \hat{z}. 
\end{equation} As before, the decoherence process on the first monomer has the effect of shrinking the length of its Bloch vector of a factor $\cos^2{\eta}$ for every qubit in the reservoir. 

Here comes the difference with the previous scenario: since now the environment we are dealing with is a Non-Markovian one, we have to \textit{store} in the reservoir qubits the information about the interaction with the first monomer of the chain. Thus, we have also to compute the states of the reservoir qubits \textit{after} the interaction with $M_1$:
\begin{equation}
    ^1\xi^{'}_1=Tr_{{^0\rho_{S_1}}}\left[\PSWAP_{{^0\rho_{S_1}},{^0\xi_1}}  {\left({^0\rho_{S_1}}\otimes{^0\xi_1}\right)} \PSWAP^\dagger_{{^0\rho_{S_1}},{^0\xi_1}}\right]= \frac{1}{2}\left(\mathbf{I}+ {^1\Vec{r}^{(1)}_1}\cdot\Vec{\sigma}\right)= \frac{1}{2}\left(\mathbf{I}-\sin^2{\eta}\ Z_1^{res}\right) 
\end{equation}
where the Bloch vector ${^1\Vec{r}^{(1)}_1}$ evolves as:
\begin{equation}
    ^1\Vec{r}^{(1)}_1=\sin^2{\eta}\ {^0\Vec{s}_1} +\cos^2{\eta}\ {^0\Vec{r}_1} +\cos{\eta}\sin{\eta}\left({^0\Vec{r}_1} \times {^0\Vec{s}_1}\right)=-\sin^2{\eta}\ \hat{z}. 
\end{equation}

Since the reservoir qubits are all, at this stage, initialised in the maximally mixed state, we can easily generalise this process to the $3^{rd}$ interaction of the monomer $M_1$ with the reservoir, to have:
\begin{equation}
    ^1\rho^{(3)}_{S_1}= \frac{1}{2}\left(\mathbf{I}+ {^1\Vec{s}^{(3)}_1}\cdot\Vec{\sigma}\right)= \frac{1}{2}\left(\mathbf{I}-\cos^{6}{\eta}\ Z_1\right)
\end{equation} where:
\begin{equation}
    ^1\Vec{s}^{(3)}_1=\cos^2{\eta}\ {^1\Vec{s}^{(2)}_1} +\sin^2{\eta}\ {^0\Vec{r}_3} -\cos{\eta}\sin{\eta}\left({^0\Vec{r}_3} \times {^0\Vec{s}^{(2)}_1}\right)=\cos^2{\eta}(-\cos^{4}{\eta}) \hat{z}=-\cos^{6}{\eta}\ \hat{z},
    \label{eq:inistatesepNONmarkov}
\end{equation} while the second and third reservoir qubit will be left in the states:
\begin{equation}
    ^1\xi^{'}_2= \frac{1}{2}\left(\mathbf{I}+ {^1\Vec{r}^{(1)}_2}\cdot\Vec{\sigma}\right)= \frac{1}{2}\left(\mathbf{I}-\sin^2{\eta}\cos^2{\eta}\ Z_2^{res}\right)
\end{equation} and
\begin{equation}
    ^1\xi^{'}_3= \frac{1}{2}\left(\mathbf{I}+ {^1\Vec{r}^{(1)}_3}\cdot\Vec{\sigma}\right)= \frac{1}{2}\left(\mathbf{I}-\sin^2{\eta}\cos^4{\eta}\ Z_3^{res}\right),
\end{equation} where:
\begin{equation}
    ^1\Vec{r}^{(1)}_2=\sin^2{\eta}\ {^1\Vec{s}^{(1)}_1} +\cos^2{\eta}\ {^0\Vec{r}_2} +\cos{\eta}\sin{\eta}\left({^0\Vec{r}_2} \times {^1\Vec{s}^{(1)}_1}\right)=-\sin^2{\eta}\cos^2{\eta}\ \hat{z} 
\end{equation} and
\begin{equation}
    ^1\Vec{r}^{(1)}_3=\sin^2{\eta}\ {^1\Vec{s}^{(2)}_1} +\cos^2{\eta}\ {^0\Vec{r}_3} +\cos{\eta}\sin{\eta}\left({^0\Vec{r}_3} \times {^1\Vec{s}^{(2)}_1}\right)=-\sin^2{\eta}\cos^4{\eta}\ \hat{z}. 
\end{equation}

Now is the turn of the second monomer $M_2$ to undergo the decoherence phase. The qubits in the reservoir are now in the states ${^1\xi^{'}_j}$, $j=1,2,3$. The process is the same described before, giving at the end of the interaction of $M_2$ with the third qubit in the reservoir:
\begin{equation}
    ^1\rho^{(3)}_{S_2}= \frac{1}{2}\left[\mathbf{I}+\cos^{4}{\eta}\left(\cos^2{\eta}-3\sin^4{\eta}\right)Z_2\right]
\end{equation} and
\begin{equation}
\begin{cases}
    ^1\xi^{''}_1= \frac{1}{2}\left[\mathbf{I}+\sin^2{\eta}\left(1-\cos^2{\eta}\right)\ Z_1^{res}\right] \\
    ^1\xi^{''}_2= \frac{1}{2}\left[\mathbf{I}+\sin^2{\eta}\left(\cos^2{\eta}-\cos^4{\eta}-\sin^4{\eta}\right)\ Z_2^{res}\right] \\
    ^1\xi^{''}_3= \frac{1}{2}\left[\mathbf{I}+\sin^2{\eta}\cos^2{\eta}\left(\cos^2{\eta}-\cos^4{\eta}-2\sin^4{\eta}\right)\ Z_3^{res}\right].
\end{cases}
\end{equation}
To conclude the decoherence phase for the $k=1$ iteration, we move now to the monomer $M_3$ considering the reservoir qubits in ${^1\xi^{''}_j}$, $j=1,2,3$:
\begin{equation}
    ^1\rho^{(3)}_{S_3}= \frac{1}{2}\left\{[\mathbf{I}+\left[3\sin^4{\eta}\cos^{2}{\eta}\left(\cos^2{\eta}-\cos^4{\eta}-\sin^4{\eta}\right)+\cos^6{\eta}\right]Z_3\right\} 
\end{equation} and
\begin{equation}
    \begin{cases}\!
        ^1\xi^{'''}_1=\frac{1}{2}\left\{\mathbf{I}+\left[\sin^2{\eta}\cos^2{\eta}\left(1-\cos^2{\eta}\right)+\sin^2{\eta}\right]Z_1^{res}\right\} \\
        ^1\xi^{'''}_2= \frac{1}{2}\left[\mathbf{I}+\frac{1}{32}\sin^2{\eta}(3\cos{2\eta}+6\cos{4\eta} -3\cos{6\eta}+26)Z_2^{res}\right] \\
        ^1\xi^{'''}_3= \frac{1}{2}\left[\mathbf{I}+\frac{1}{64}\sin^2{\eta}(58\cos{2\eta}-8\cos{4\eta} +6\cos{6\eta}-3\cos{8\eta}+11)Z_3^{res}\right].
    \end{cases}
    \label{eq:inicondk2NONmarkov}
\end{equation}

To conclude the first iteration of the protocol we move now to the \textit{transfer phase} involving the monomers $M_1$ and $M_2$. The unitary describing this process is the one we introduced in Eq.\ref{eq:unitary}, so that:
\begin{equation}
    ^1\rho_{S_{1,2}}=U\left({^1\rho^{(3)}_{S_1}}\otimes{^1\rho^{(3)}_{S_2}}\right)U^\dagger
\end{equation} which gives, at first order in $t$:
\begin{equation}
    ^1\rho_{S_{1,2}}={^1\rho^{(3)}_{S_1}}\otimes{^1\rho^{(3)}_{S_2}}+itJ_1\cos^4{\eta}\left(\cos^2{\eta}-\frac{3}{2}\sin^4{\eta}\right)\left(\sigma^+_1\sigma^-_2-\sigma^-_1\sigma^+_2\right).
    \label{eq:inicondk2NONmarkovstate}
\end{equation}
This concludes the first iteration of our protocol, leaving the polymer in the state:
\begin{equation}
    ^2\rho_S={^1\rho_{S_{1,2}}}\otimes{^1\rho^{(3)}_{S_3}}
\end{equation} and the reservoir qubits in the states in Eq.\ref{eq:inicondk2NONmarkov}.

\textit{Second iteration $k=2$ and recombination}.--
We move now to the second and final iteration of the protocol before the recombination of the exciton and, differently from the Markov scenario, the initial conditions for the reservoir will be given by the states in Eq.\ref{eq:inicondk2NONmarkov}, as we want the environment to keep \textit{memory} of the previous steps of the process. This brings in some differences in the way in which separable and non-separable parts of Eq.\ref{eq:inicondk2NONmarkovstate} evolve, which we will attention now.

Let us start with the first monomer $M_1$, which interacts with the first qubit in the reservoir with the usual \PSWAP:
\begin{align}
   ^2\rho^{(1)}_{S_{1,2}}=&Tr_{{^2\xi^{'''}_1}}\left[\PSWAP\left({^1\rho_{S_{1,2}}}\otimes{^1\xi^{'''}_1}\right)\PSWAP^\dagger\right] & \\ = & Tr_{{^2\xi^{''}_1}}\left[\PSWAP\left({^1\rho^{(3)}_{S_1}}\otimes{^1\rho^{(3)}_{S_2}}\otimes{^1\xi^{'''}_1}\right)\PSWAP^{\dagger}\right] \label{eq:k2sepNONmarkov}  & \\
    & + itJ_1\cos^4{\eta}\left(\cos^2{\eta}-\frac{3}{2}\sin^4{\eta}\right) Tr_{{^2\xi^{''}_1}}\left[\PSWAP\left(\left(\sigma^+_1\sigma^-_2-\sigma^-_1\sigma^+_2\right)\otimes{^1\xi^{'''}_1}\right)\PSWAP^{\dagger}\right]
    \label{eq:k2NONsepNONmarkov}
\end{align}
The separable part in Eq.\ref{eq:k2sepNONmarkov} evolves as explained before, with the only difference that now the initial conditions will be given by $^1\Vec{s}^{(3)}_1$ in Eq.\ref{eq:inistatesepNONmarkov} and $^1\Vec{r}^{(3)}_1=\sin^2{\eta}\cos^2{\eta}(1-\cos^2{\eta})+\sin^2{\eta}$, so to have:
\begin{equation}
   {^2\rho^{(1)}_{S_1}}\otimes{^2\rho^{(0)}_{S_2}}=\frac{1}{2}\left\{\textbf{I}+{^2s^{(1)}_1(\eta)}Z_1\right\}\otimes{^2\rho^{(0)}_{S_2}},
\end{equation}
where:
\begin{equation}
    {^2s^{(1)}_1(\eta)}=-\left[\cos^8{\eta}+\frac{1}{8}\sin^4{\eta}\left(\cos{4\eta-9}\right)\right]
\end{equation} 
is the $\hat{z}$ component of the Bloch vector ${^2\Vec{s}_1}$ after the interaction with the first reservoir qubit.

Moving now to the non separable part in Eq.\ref{eq:k2NONsepNONmarkov}, that we will call here $^1\rho^{non-sep}_{S_{1,2}}$, we have:
\begin{align}
    ^1\rho^{non-sep,(1)}_{S_{1,2}}=&itJ_1\cos^4{\eta}\left(\cos^2{\eta}-\frac{3}{2}\sin^4{\eta}\right)\left[\cos^2{\eta}\left(\sigma^+_1\sigma^-_2-\sigma^-_1\sigma^+_2\right) \right.&  \\
    & \left. +i\sin^3{\eta}\cos{\eta}\left(\cos^2{\eta}-\cos^4{\eta}+1\right)\left(\sigma^+_1\sigma^-_2+\sigma^-_1\sigma^+_2\right) \right], 
\end{align} that we can conveniently write as:
\begin{align}
    ^1\rho^{non-sep,(1)}_{S_{1,2}}=&itJ_1\left[F^{(1,0)}(\eta)\left(\sigma^+_1\sigma^-_2-\sigma^-_1\sigma^+_2\right) \right. \label{eq:commonhop} & \\ & \left. + G^{(1,0)}(\eta)\left(\sigma^+_1\sigma^-_2+\sigma^-_1\sigma^+_2\right)\right] \label{eq:sepNONmark}
\end{align} where:
\begin{equation}
    \begin{cases}
        F^{(1,0)}(\eta)=\cos^6{\eta}\left(\cos^2{\eta}-\frac{3}{2}\sin^4{\eta}\right) \\
        G^{(1,0)}(\eta)=i\cos^4{\eta}\sin^5{\eta}\left(\cos^2{\eta}-\frac{3}{2}\sin^4{\eta}\right)\left(\cos^2{\eta}-\cos^4{\eta}+1\right).
    \end{cases}
\end{equation}
The superscript $(l,m)$ indicates the stage of the interactions of the two entangled monomers with the reservoir: the first label refers to $M_1$, the second to $M_2$. Comparing this result with Eq.\ref{eq:nonsepMark}, we notice that a new hopping term appears, in Eq.\ref{eq:sepNONmark}, which shows a $\pi$-phase difference with the one in Eq.\ref{eq:commonhop}: it comes from the coherence of the reservoir qubit due to its non-markovianity. Interestingly, the phase factor would have been different depending on the orientation of the reservoir qubit's Bloch vector $\Vec{r}_1$ on the Bloch sphere: in our model,  it is a $\pi$-phase because ${^1\Vec{r}_1^{(3)}}$ is parallel to the $\hat{z}$-axis of the sphere. \\
Before moving on, we have to update the state of the first reservoir qubit, because of the non-markovianity of the environment we are dealing with:
\begin{equation}
    ^2\xi_1^{'}=\frac{1}{2}\left[\mathbf{I}+{^2r^{'}_1}(\eta)Z^{res}_1\right]
\end{equation}
where:
\begin{equation}
   ^2r^{'}_1(\eta)=\cos^2{\eta}\sin^4{\eta}\left(2+\cos{2\eta}\right).    
\end{equation}
Here $^2r^{'}_1(\eta)$ is the $\hat{z}$ component of the Bloch vector ${^2\Vec{r}_1}$ after the interaction with the monomer $M_1$.

We can now proceed with the second monomer $M_2$ and its interaction with the first qubit in the reservoir. Splitting again separable and non-separable parts as before, we find:
\begin{equation}
   {^2\rho^{(1)}_{S_1}}\otimes{^2\rho^{(1)}_{S_2}}={^2\rho^{(1)}_{S_1}}\otimes\frac{1}{2}\left[\textbf{I}+{^2s^{(1)}_2} Z_2\right],
\end{equation} where
\begin{equation}
    ^2s^{(1)}_2=\cos^2{\eta}\left(\cos^{10}{\eta}+\sin^4{\eta}+\cos^2{\eta}\sin^4{\eta}-5\cos^4{\eta}\sin^4{\eta}\right),
\end{equation}
and:
\begin{align}
    ^1\rho^{non-sep,(2)}_{S_{1,2}}=&itJ_1\left[F^{(1,1)}(\eta)\left(\sigma^+_1\sigma^-_2-\sigma^-_1\sigma^+_2\right) \right.  & \\ & \left. + G^{(1,1)}(\eta)\left(\sigma^+_1\sigma^-_2+\sigma^-_1\sigma^+_2\right)\right] 
\end{align} being:
\begin{equation}
    \begin{cases}
        F^{(1,1)}(\eta)=\cos^2{\eta}F^{(1,0)}(\eta)+i\sin{\eta}\cos{\eta}G^{(1,0)}(\eta){^2r^{'}_1}(\eta)\\
        G^{(1,1)}(\eta)=\cos^2{\eta}G^{(1,0)}(\eta)-i\sin{\eta}\cos{\eta}F^{(1,0)}(\eta){^2r^{'}_1}(\eta).
    \end{cases}
\end{equation} 
Before moving on with the third monomer $M_3$, we have to extract the new state of the first reservoir qubit:
\begin{equation}
    ^2\xi_1^{''}=\frac{1}{2}\left[\mathbf{I}+{^2r^{''}_1}(\eta)Z_1^{res}\right]
\end{equation} where:
\begin{equation}
    ^2r^{''}_1(\eta)=\cos^4{\eta}\sin^2{\eta}\left(\cos^2{\eta}+\cos^6{\eta}-5\sin^4{\eta}+1\right).
\end{equation}

We can thus conclude the interactions with the first reservoir qubit considering the monomer $M_3$. At the $k=2$ iteration of the protocol, it has not been involved in the transfer phase yet, meaning that it is not entangled with the monomers $M_1$ and $M_2$. We simply have to write then:
\begin{equation}
    ^2\rho^{(1)}_{S_3}=Tr_{{^2\xi^{''}_1}}\left[\PSWAP\left({^1\rho^{(3)}_{S_3}}\otimes{^2\xi^{''}_1}\right)\PSWAP^\dagger\right]=\frac{1}{2}\left\{\textbf{I}+{^2s^{(1)}_3}Z_3\right\}, 
\end{equation}
with:
\begin{equation}
    ^2s^{(1)}_3=\left[\cos^8{\eta}+\sin^4{\eta}\cos^4{\eta}\left(4\cos^2{\eta}-3\cos^4{\eta}+\cos^6{\eta}-8\sin^4{\eta}+1\right)\right],
\end{equation} and:
\begin{equation}
    ^2\xi^{'''}_1=Tr_{{^1\rho^{(3)}_{S_3}}}\left[\PSWAP\left({^1\rho^{(3)}_{S_3}}\otimes{^2\xi^{''}_1}\right)\PSWAP^\dagger\right]=\frac{1}{2}\left[\textbf{I}+{^2r^{'''}_1}Z_1^{res}\right],
\end{equation} where:
\begin{equation}
    ^2r^{'''}_1=-\frac{1}{64}\cos^2{\eta}\sin^2{\eta}\left(25-332\cos{2\eta}+68\cos{4\eta}-20\cos{6\eta}+3\cos{8\eta}\right)
\end{equation}

We have now to move to the second and third reservoir qubits in our Non-Markov environment going through the very same steps outlined above. The state of the polymer at end of the decoherence phase of the $k=2$ iteration of the protocol in the Non-Markov scenario will then be:
\begin{equation}
    ^2\rho^{(3)}_{S}={^2\rho^{(3)}_{S_{1,2}}}\otimes{^2\rho^{(3)}_{S_3}}
    \label{eq:finalk2iterNONmark}
\end{equation} where:
\begin{equation}
    {^2\rho^{(3)}_{S_{1,2}}}={^2\rho^{(3)}_{S_1}}\otimes{^2\rho^{(3)}_{S_2}}+itJ_1\left[F(\eta)\left(\sigma^+_1\sigma^-_2-\sigma^-_1\sigma^+_2\right)+ G(\eta)\left(\sigma^+_1\sigma^-_2+\sigma^-_1\sigma^+_2\right)\right]
\end{equation} with:
\begin{equation}
    ^2\rho^{(3)}_{S_1}=\frac{1}{2}\left[\textbf{I}+{^2s^{(3)}_1}Z_1\right],
\end{equation} 
\begin{equation}
    ^2s^{(3)}_1=\frac{1}{2048}\left(1302-128\cos{2\eta}+655\cos{4\eta} +128\cos{6\eta}+90\cos{8\eta}+\cos{12\eta}\right);
\end{equation}
\begin{equation}
    ^2\rho^{(3)}_{S_2}=\frac{1}{2}\left[\textbf{I}+{^2s^{(3)}_2}Z_2\right],
\end{equation}
\begin{equation}
    ^2s^{(3)}_2=\left[\cos^{16}{\eta}+\sin^{12}{\eta}-\frac{\cos^2{\eta}\sin^4{\eta}}{128}\left(-147-18\cos{2\eta}+12\cos{4\eta}+17\cos{6\eta}+7\cos{8\eta}+\cos{10\eta}\right)\right],
\end{equation}
and:
\begin{align}
    &F(\eta)= F^{(3,3)}(\eta)=\cos^2{\eta}F^{(3,2)}(\eta)+i\sin{\eta}\cos{\eta}G^{(3,2)}(\eta){^2r^{'}_3} \nonumber & \\ 
    &=-\frac{1}{2.48\cdot10^{27}}\left[\cos^{14}{\eta}\left(1.84\cdot10^{27}\cos{2\eta}-2.62\cdot10^{27}\cos{4\eta}+9.33\cdot10^{26}\cos{6\eta}-8.97\cdot10^{25}\cos{8\eta}-1.34\cdot10^{26}\cos{10\eta}   \nonumber \right. \right. & \\ & \left. \left. 
    +1.63\cdot10^{26}\cos{12\eta}-1.22\cdot10^{26}\cos{14\eta}+6.98\cdot10^{25}\cos{16\eta}-3.50\cdot10^{25}\cos{18\eta}+1.54\cdot10^{25}\cos{20\eta}-6.13\cdot10^{24}\cos{22\eta}  \nonumber \right. \right. & \\ & \left. \left. +2.23\cdot10^{24}\cos{24\eta}-7.00\cdot10^{23}\cos{26\eta}+1.72\cdot10^{23}\cos{28\eta}
    -2.23\cdot10^{22}\cos{30\eta}-7.94\cdot10^{21}\cos{32\eta}
    +8.59\cdot10^{21}\cos{34\eta}  \nonumber \right. \right. & \\ & \left. \left. -4.83\cdot10^{21}\cos{36\eta}
    +2.14\cdot10^{21}\cos{38\eta}-8.00\cdot10^{20}\cos{40\eta}
    +2.53\cdot10^{20}\cos{42\eta}-6.56\cdot10^{19}\cos{44\eta}
    +1.16\cdot10^{19}\cos{46\eta}  \nonumber \right. \right. & \\ & \left. \left. +3.15\cdot10^{17}\cos{48\eta}-1.42\cdot10^{18}\cos{50\eta}+7.92\cdot10^{17}\cos{52\eta}
    -3.05\cdot10^{17}\cos{54\eta}+8.99\cdot10^{16}\cos{56\eta}
    -1.94\cdot10^{16}\cos{58\eta}   \nonumber \right. \right. & \\ & \left. \left. +1.99\cdot10^{15}\cos{60\eta}
    +6.90\cdot10^{14}\cos{62\eta}-5.02\cdot10^{14}\cos{64\eta}+1.85\cdot10^{14}\cos{66\eta}
    -4.80\cdot10^{13}\cos{68\eta}+7.93\cdot10^{12}\cos{70\eta}  \nonumber \right. \right. & \\ & \left. \left. +2.69\cdot10^{10}\cos{72\eta}
    -5.41\cdot10^{11}\cos{74\eta}+1.90\cdot10^{11}\cos{76\eta}-3.09\cdot10^{10}\cos{78\eta}
    +5.51\cdot10^{8}\cos{80\eta}+7.30\cdot10^{8}\cos{82\eta}  \nonumber \right. \right. & \\ & \left. \left. -1.18\cdot10^{8}\cos{84\eta}+2.79\cdot10^{6}\cos{86\eta}+9.61\cdot10^{5}\cos{88\eta}-7.31\cdot10^{4}\cos{90\eta}-1.78\cdot10^{3}\cos{92\eta}+243\cos{94\eta}-2.49\cdot10^{27}\right)\right] 
\end{align}
\begin{align}
    &G(\eta)= G^{(3,3)}(\eta)=\cos^2{\eta}G^{(3,2)}(\eta)-i\sin{\eta}\cos{\eta}F^{(3,2)}(\eta){^2r^{'}_3} \nonumber & \\ 
    &=\frac{i\sin^3{\eta}\cos^{13}{\eta}}{1.51\cdot10^{23}}\left[-5.43\cdot10^{22}\cos{2\eta}+1.01\cdot10^{23} \cos{4\eta}+8.20\cdot10^{21}\cos{6\eta}+2.12\cdot10^{22}\cos{8\eta}+5.32\cdot10^{21}\cos{10\eta}    \nonumber \right. & \\ &  \left. -3.36\cdot10^{21}\cos{12\eta}+2.26\cdot10^{21}\cos{14\eta}-1.53\cdot10^{21}\cos{16\eta}+7.24\cdot10^{20}\cos{18\eta}-1.96\cdot10^{20}\cos{20\eta}+3.24\cdot10^{19}\cos{22\eta}     \nonumber \right. & \\ &  \left. +1.64\cdot10^{19}\cos{24\eta}-1.83\cdot10^{19}\cos{26 \eta}+1.09\cdot10^{19}\cos{28\eta}-5.12\cdot10^{18}\cos{30\eta}+1.88\cdot10^{18}\cos{32\eta}-5.17\cdot10^{17}\cos{34\eta}     \nonumber \right. & \\ &  \left. +9.20\cdot10^{16}\cos{36\eta}+2.65\cdot10^{15}\cos{38\eta}-1.24\cdot10^{16}\cos{40\eta}+7.07\cdot10^{15}\cos{42\eta}-2.61\cdot10^{15}\cos{44\eta}+6.79\cdot10^{14}\cos{46\eta}     \nonumber \right. & \\ &  \left. -1.02\cdot10^{14}\cos{48\eta}-6.60\cdot10^{12}\cos{50\eta}+1.12\cdot10^{13}\cos{52\eta}-4.94\cdot10^{12}\cos{54\eta}+1.50\cdot10^{12}\cos{56\eta}-2.95\cdot10^{11}\cos{58\eta}     \nonumber \right. & \\ &  \left. +7.72\cdot10^{9}\cos{60\eta}+1.83\cdot10^{10}\cos{62\eta}-6.35\cdot10^{9}\cos{64\eta}+8.40\cdot10^{8}\cos{66\eta}+3.30\cdot10^{7}\cos{68\eta}-2.53\cdot10^{7}\cos{70\eta}     \nonumber \right. & \\ &  \left. +2.84\cdot10^{6}\cos{72\eta}-3.28\cdot10^{4}\cos{74\eta}-1.41\cdot10^{4}\cos{76\eta}+729\cos{78\eta}+7.17\cdot10^{22}\right]
\end{align}
while:
\begin{equation}
    ^2\rho^{(3)}_{S_3}=\frac{1}{2}\left[\textbf{I}+s(\eta)Z_3\right], 
\end{equation} $s(\eta)$ being:
\begin{align}
    s(\eta)= {^2s^{(3)}_3}=&\cos^{12}{\eta}+\sin^{12}{\eta}+\sin^4{\eta}\left(3\cos^8{\eta}+6\cos^{10}{\eta}-3\cos^{12}{\eta}+4\cos^{14}{\eta}\right)+\sin^8{\eta}\left(6\cos^4{\eta}+7\cos^{6}{\eta}-27\cos^{8}{\eta}\right) & \nonumber \\ & +\sin^{12}{\eta}\left(-\cos^2{\eta}-2\cos^{4}{\eta}-4\cos^{6}{\eta}\right). 
\end{align}
\begin{figure}
    \centering
    \includegraphics[width=0.4\linewidth]{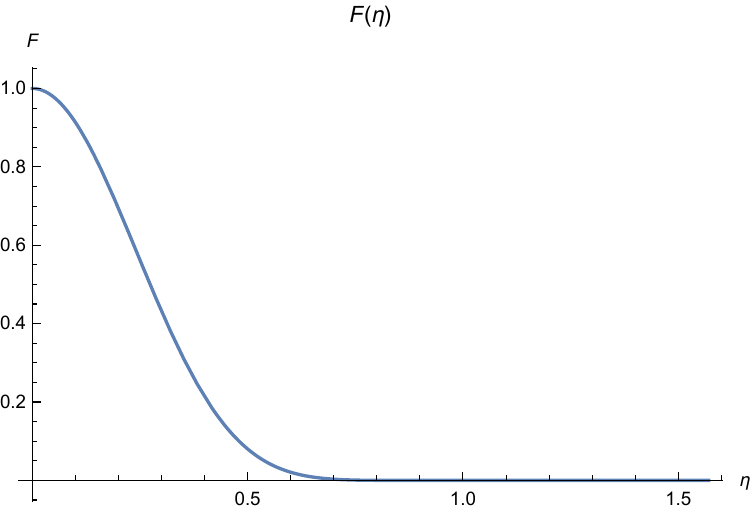} \hfill
    \includegraphics[width=0.4\linewidth]{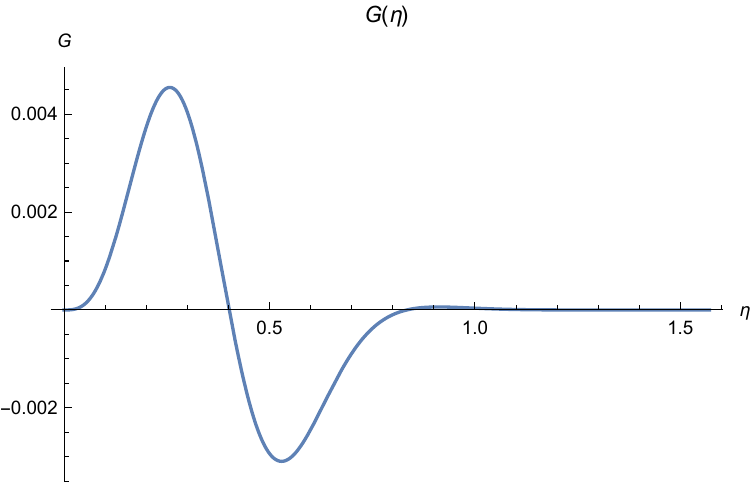} \hfill
    \includegraphics[width=0.4\linewidth]{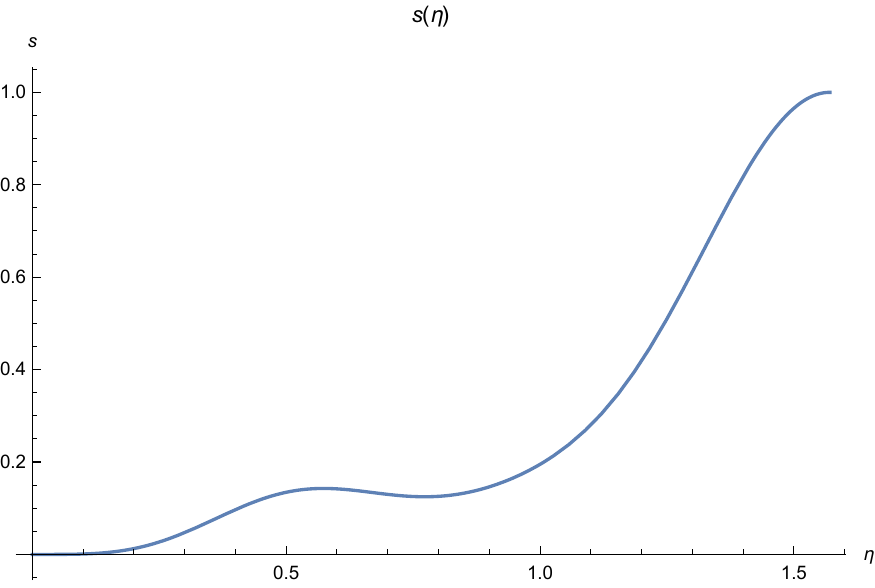}
    \caption{Coefficient $F(\eta)$, $G(\eta)$ and $s(\eta)$ at the end of $k=2$ iteration.}
    \label{fig:coeffvseta}
\end{figure}
These complex coefficients are shown as a function of the decoherence parameter $\eta$ in Fig\ref{fig:coeffvseta}.

Once the decoherence phase ends, the transfer phase of the $k=2$ iteration of the protocol can take place. We thus apply again the unitary in Eq.\ref{eq:unitary} to the state $^2\rho^{(3)}_S$ in Eq.\ref{eq:finalk2iterNONmark} to obtain:
\begin{align}
    ^3\rho_S=&{^3\rho_{S_1}}\otimes{^3\rho_{S_2}}\otimes{^3\rho_{S_3}} + itJ_1\left[F(\eta)+2itB_2G(\eta)\right]\left(\sigma^+_1\sigma^-_2 - \sigma^-_1\sigma^+_2\right)\otimes{^3\rho_{S_3}} \nonumber  & \\ 
    &  +itJ_1\left[G(\eta)+2itB_2F(\eta)\right] \left(\sigma^+_1\sigma^-_2 + \sigma^-_1\sigma^+_2\right)\otimes{^3\rho_{S_3}} \nonumber & \\
    & -\frac{1}{2}t^2J_1J_2\left(Z_2+s(\eta)\right)\left[F(\eta)\left(\sigma^+_1\sigma^-_3 + \sigma^-_1\sigma^+_3\right)+G(\eta)\left(\sigma^+_1\sigma^-_3 - \sigma^-_1\sigma^+_3\right)\right]
\end{align} so that, after the recombination of the exciton, the emitted photon is in the state in Eq.\ref{eq:finalstateNonMarkov}. 

\newpage
\section{Comparison between $F(\eta)$, $G(\eta)$ and the Markovian equivalent for the exciton transfer \textit{via the chain}}

In this Appendix, we include the tables detailing the values of the coefficients of the terms describing the coherent delocalisation of the exciton \textit{via the chain} for different values of the decoherence parameter $\eta$. The goal of this Appendix is to support the claims that in the weak coupling regime (1) $F(\eta)$ and its Markovian counterpart have the same value \textit{at every step} of the protocol and (2) that $G(\eta)$ is small enough to be negligible. 

\begin{table}[hb]
    \centering
        \begin{tabular}{|c|c|c|c|}
    \hline
                              $\eta=0.01$      & Markov &   $F(\eta)$ & $G(\eta)$ \\
    \hline
        $M_1\leftrightarrow Q_1$&   0.9996    &     0.9996   &  $i9.9\cdot10^{-7}$          \\           
        $M_1\leftrightarrow Q_2$&   0.9995    &     0.9995  &  $i9.9\cdot10^{-7}$          \\
        $M_1\leftrightarrow Q_3$&   0.9994    &     0.9994   &  $i1.9\cdot10^{-6}$          \\
        $M_2\leftrightarrow Q_1$&   0.9993    &     0.9993   &  $i3.9\cdot10^{-10}$          \\
        $M_2\leftrightarrow Q_2$&   0.9992    &     0.9992   &  $i9.9\cdot10^{-7}$          \\
        $M_2\leftrightarrow Q_3$&   0.9991    &     0.9991   &  $-i9.9\cdot10^{-7}$        \\  
    \hline
    \end{tabular}
    \begin{tabular}{|c|c|c|c|}
    \hline
                             $\eta=0.1$       & Markov &   $F(\eta)$ & $G(\eta)$ \\
    \hline
        $M_1\leftrightarrow Q_1$&   0.961    &     0.961   &  $i9.7\cdot10^{-4}$          \\           
        $M_1\leftrightarrow Q_2$&   0.951    &     0.951   &  $i9.3\cdot10^{-4}$          \\
        $M_1\leftrightarrow Q_3$&   0.942    &     0.942   &  $i1.3\cdot10^{-3}$          \\
        $M_2\leftrightarrow Q_1$&   0.932    &     0.932   &  $i3.7\cdot10^{-5}$          \\
        $M_2\leftrightarrow Q_2$&   0.923    &     0.923   &  $i9.4\cdot10^{-4}$          \\
        $M_2\leftrightarrow Q_3$&   0.914    &     0.914   &  $-i8.3\cdot10^{-4}$        \\  
    \hline
    \end{tabular}
    \begin{tabular}{|c|c|c|c|}
    \hline
                              $\eta=1.0$      & Markov &   $F(\eta)$ & $G(\eta)$ \\
    \hline
        $M_1\leftrightarrow Q_1$&   $7.263\cdot10^{-3}$   &     $-1.173\cdot10^{-2}$   &  $-i1.52\cdot10^{-2}$          \\           
        $M_1\leftrightarrow Q_2$&   $2.120\cdot10^{-3}$    &     $-1.736\cdot10^{-3}$  &  $-i3.2\cdot10^{-3}$          \\
        $M_1\leftrightarrow Q_3$&   $6.189\cdot10^{-4}$    &     $7.826\cdot10^{-5}$   &  $-i9.3\cdot10^{-4}$          \\
        $M_2\leftrightarrow Q_1$&   $1.807\cdot10^{-4}$    &     $2.117\cdot10^{-4}$   &  $-i3.1\cdot10^{-4}$          \\
        $M_2\leftrightarrow Q_2$&   $5.275\cdot10^{-5}$    &     $7.715\cdot10^{-5}$   &  $-i8.8\cdot10^{-5}$          \\
        $M_2\leftrightarrow Q_3$&   $1.540\cdot10^{-5}$    &     $3.501\cdot10^{-5}$   &  $-i3.6\cdot10^{-5}$        \\  
    \hline
    \end{tabular}
    \caption{Coefficients hopping terms via chain with $\eta=0.01$, $\eta=0.1$ and $\eta=1.0$ at every step of the decoherence phase of $k=2$ iteration.}
    \label{tab:coeff}
\end{table}

Table \ref{tab:coeff} shows the coefficients for $\eta=0.01$, $\eta=0.1$ and $\eta=1.0$. One can see that in the weak coupling regime, claims (1) and (2) are satisfied, but as soon as we increase the decoherence parameter the coefficients $F(\eta)$ and their Markovian counterpart become dissimilar. In contrast, $G(\eta)$ cannot be neglected anymore as it has the same order of magnitude as $F(\eta)$. We leave the reader to the main text for an interpretation of these results.

%% file: main.bbl
\begin{thebibliography}{46}
\bibitem [{sha()}]{sharp}%
  \BibitemOpen
  \href@noop {} {}\bibinfo {note} {In quantum physics, a variable is ``sharp" when the state of the system it belongs to is in an eigenstate of that variable. Here, $Z_Q$ is sharp on $Q$ if the state of $Q$ is either $\ket{0}$ or $\ket{1}$. For a more formal definition, see [30].}\BibitemShut {Stop}%
\end{thebibliography}
